# From Framework to Practice: Youth Negotiations of Privacy with Smart Voice Assistants Through the PEA-AI Lens

Youth Privacy in Voice Assistants

Molly Campbell

Computer Science, Vancouver Island University, Nanaimo, Canada, molly.campbell@viu.ca

Yulia Bobkova

Computer Science, Vancouver Island University, Nanaimo, Canada, yulia.bobkova@viu.ca

Ajay Kumar Shrestha

Computer Science, Vancouver Island University, Nanaimo, Canada, ajay.shrestha@viu.ca

Smart voice assistants (SVAs) have become embedded in the daily lives of youth, introducing complex privacy challenges due to always-on listening, shared device usage, and opaque data practices. This study applies the Privacy-Ethics Alignment in AI (PEA-AI) framework to examine how youth perceive and negotiate privacy within SVAs. Through a survey of 469 Canadian youth (aged 16-24), we measured five privacy constructs: Perceived Privacy Risk (PPR), Perceived Privacy Benefits (PPBf), Algorithmic Transparency and Trust (ATT), Privacy Self-Efficacy (PSE), and Privacy-Protective Behavior (PPB). Results reveal a persistent privacy paradox. While youth express moderate to high privacy concerns (PPR M = 3.61), perceived benefits (PPBf M = 3.00) and protective actions (PPB M = 3.03) remain moderate, with transparency and trust scoring lowest (ATT M = 2.52). Heavy SVA users report higher benefits and lower risk perception than light users. High protective behavior is strongly associated with both high risk perception and high self-efficacy. Qualitative insights from prior focus groups contextualize these patterns, illustrating how youth navigate tensions between convenience, control, and trust. The findings provide actionable design principles for SVA along with implications for multi-stakeholder governance and youth-centered digital literacy.

**ACM Reference Format:**
Molly Campbell, Yulia Bobkova, and Ajay K Shrestha. From Framework to Practice: Youth Negotiations of Privacy with Smart Voice Assistants Through the PEA-AI Lens: ACM J. Responsible Computing. NOTE: This block will be automatically generated when manuscripts are processed after acceptance.

## 1 INTRODUCTION

Smart voice assistants (SVAs) are becoming routine companions in the everyday lives of youth. Their use spans private bedrooms, shared living spaces, and personal mobile devices, enabling quick and hands-free access to information and media [39]. In many households, SVAs are situated on shared devices such as smart speakers, smart TVs, or tablets located in common spaces, where multiple users interact with a single assistant [29]. While convenient, this arrangement also subjects youth to passive listening and data collection. Privacy settings are commonly configured by a single adult account holder, leaving young users with reduced agency and visibility into how their data is handled [3, 11].

The dual promise of SVAs lies in the tension between convenience and persistent listening. Their hands-free capabilities support quick commands and simple integration into daily routines. The same features, however, depend on continuous voice capture and a complex data infrastructure that remains hidden from youth, limiting their capacity to understand, consent to, or influence how their speech data is processed [13, 38].

Youth privacy is difficult to manage due to a variety of factors, including complex settings, shared devices and power asymmetries with platforms. Settings are often buried within multi-layered menus, contain confusing language, and are spread across different settings and account dashboards tied to an adult's primary account, creating substantial friction for youth attempting to understand or adjust privacy controls [11, 17]. Existing work often treats youth as "users with concerns" but lacks detailed patterns of how they see risks/benefits, trust, capability, and actual behavior in the SVA context [6, 35].

In response to these challenges, the 2025-developed Privacy-Ethics Alignment in AI (PEA-AI) model contextualizes youth privacy as an ongoing negotiation among three key stakeholder groups: youth, parents and educators, and AI professionals [6]. The framework emphasizes how each stakeholder carries distinct expectations, concerns, and responsibilities regarding data practices and consent. Earlier phases include a PEA-AI journal paper and a qualitative SVA focus-group paper [6, 11].

This article focuses specifically on youth and their interactions with smart voice assistants. It introduces a measurement instrument aligned with the PEA-AI framework and examines how youth navigate privacy among five constructs and their items. It also combines qualitative focus-group narratives from earlier phases to understand how privacy expectations, perceptions, and behaviors emerge in SVA use. This study is guided by the following three research questions:

- RQ1: What are the construct-level and item-level patterns of youth perceptions and behaviors with SVAs (PPR, PPBf, ATT, PSE, PPB)?
- RQ2: How do these patterns, together with focus-group evidence, reveal typical ways youth negotiate risks, benefits, transparency, and control with SVAs?
- RQ3: What design and governance implications follow from these patterns within the PEA-AI framework?

This article contributes a PEA-AI-aligned instrument for examining youth privacy in SVA contexts. It offers a mixed-methods interpretation of how PEA-AI tensions materialize in everyday SVA use. Finally, it offers concrete and multi-level recommendations for platforms, educators, and regulators.

## 2 BACKGROUND AND RELATED WORKS

### 2.1 Youth Privacy and AI

Research on youth privacy reveals a complex landscape shaped by concerns, the privacy paradox, and varying levels of digital privacy. Young people, who are often early adopters and frequent users of technology, frequently articulate anxieties about data collection, surveillance, and a lack of control over their personal information [2, 42]. However, these concerns often do not translate into protective behaviors, a phenomenon known as the privacy paradox[2, 5, 37]. This gap is frequently attributed to the appeal of convenience, social pressure, and interface design that obscures privacy controls [28]. Furthermore, digital literacy and privacy self-efficacy, which is a youth's belief in their ability to manage privacy settings, are important but unevenly distributed, influencing how risks are perceived and navigated [24, 33].

These challenges are particularly pronounced in AI-driven environments, where young digital citizens engage heavily with personalized and algorithmic settings [35]. Despite their familiarity with technology, many lack an established understanding of data ownership, long-term digital footprints, and data collection practices of AI systems [16]. The opaque data practices, combined with the immediate benefits of personalization and social connectivity, often lead youth to prioritize short-term gains over long-term privacy risks, reinforcing the privacy paradox in AI contexts [21, 41]. This lack of transparency and trust results in youth frequently reporting uncertainty about how their data is collected and used, which erodes trust and discourages informed privacy decision-making [14].

SVAs represent a distinct case within this broader context due to several unique attributes. First, their always-on architecture, which is necessary for wake-word detection, creates a sense of ambient surveillance and raises questions about what is recorded, when, and for how long [13, 38]. Second, SVAs have become deeply embedded in private spaces, often operating as shared household devices [29, 39]. This introduces dynamics of parental mediation and shared control, where privacy settings may be managed by parents, potentially limiting youth autonomy and creating an "efficacy gap", especially for younger adolescents (16-18) who report higher trust but lower translation of trust into protective actions [3, 8]. Third, emerging evidence suggests that gender moderates key privacy pathways, with male youth showing a stronger direct link from perceived risk to action, while female youth rely more on trust and self-efficacy as a precursor to action [10]. Finally, the opaque data flows seen in the AI ecosystem complicate youth understanding of data lifecycles, undermining transparency and trust [17, 22]. These findings highlight the need for inclusive, developmentally sensitive, and household-aware privacy designs that account for the varied social, cognitive, and contextual factors shaping youth privacy experiences with SVAs.

### 2.2 The PEA-AI Framework

The PEA-AI framework provides a structured, stakeholder-centric model for providing understanding and governing privacy in AI-driven ecosystems [6]. Developed through a grounded theory analysis of perspectives from young digital citizens, AI professionals, and parents/educators, the framework conceptualizes privacy not as a fixed set of rules, but as a dynamic process of negotiations among stakeholders with often differing priorities and expectations. It proposes five core dimensions that collectively shape privacy perceptions and behaviors in AI landscapes.

The first dimension, Data Ownership and Control (DOC), examines the degree to which individuals, particularly youth, feel they can exercise control and autonomy over their personal data within AI systems. These dimensions highlight the tensions between users' desire for self-efficacy and the technical and structural limitations of current AI architectures. The second dimension, Perceived Risks and Benefits (PRB), captures the trade-offs stakeholders make when engaging with AI, weighing potential privacy harms, such as data misuse and surveillance, against perceived advantages like personalization and efficiency. This calculus is central to the well-documented "privacy paradox" observed in user behavior [19, 28]. The third dimension, Transparency and Trust (TT), focuses on how openness and explainability in AI design and data practices play a foundational role in building user trust. Furthermore, the framework incorporates Education and Awareness (EA) as a key dimension, which assesses the impact of AI literacy and privacy education on stakeholders' ability to navigate AI environments. This dimension highlights the significant knowledge gap that hinders effective privacy management and self-efficacy. Finally, the Parental Data Sharing (PDS) dimension investigates the complex role of parents and educators in mediating youth privacy. It reveals fundamental tensions between protective oversight and youth autonomy, sparking questions on when and how adult intervention is appropriate in data-related decisions.

The PEA-AI focuses on how these dimensions interact in a negotiation-based view of privacy governance. Stakeholders like youth, parents/educators, and AI professionals bring different values and knowledge to this negotiation, leading to distinct perspectives on each dimension. Key tensions emerge from these differences, including conflicts between risk and benefit, control and capability, and transparency and trust. By framing privacy as an ongoing, multi-stakeholder negotiation, the PEA-AI model moves beyond one-size-fits-all solutions. It provides a foundation for developing adaptive, inclusive, and ethical AI governance strategies that are responsive to the lived experiences and evolving expectations of its users, with an emphasis on protecting and empowering young digital citizens.

### 2.3 Constructs for Youth Privacy

Our study of youth privacy behavior in SVA ecosystems is centered around five core constructs derived from empirical research: Perceived Privacy Risk (PPR), Perceived Privacy Benefits (PPBf), Algorithmic Transparency and Trust (ATT). Privacy Self-Efficacy (PSE), and Privacy-Protective Behavior (PPB). These constructs form a cohesive framework that conceptualizes youths' attitudes and behaviors regarding privacy in SVA systems.

The construct of PPR and PPBf directly quantifies the Privacy Calculus framework, which suggests that an individual's privacy decision-making is a trade-off between perceived risks and perceived benefits [2, 5], and aligns with the PRB dimension of the PEA-AI framework. In the SVA context, PPR captures youth concerns about continuous data capture and potential issues [26, 36], while PPBf represents the perceived value derived from convenience and personalization [1, 15]. The ATT construct connects the critical link between a system's perceived openness and user trust [7, 20]. It measures youth's perception that SVA developers are transparent about data practices, which in turn fosters a sense of trust in these systems, directly aligning with the Transparency and Trust (TT) dimension.

PSE maps to two PEA-AI dimensions: Data Ownership and Control (DOC) and Education and Awareness (EA). It represents the psychological capability and confidence to exercise control over one's data, aligning with the core of DOC as conceptualized in youth-centric models [9, 12, 34]. Furthermore, this capability is dependent on the knowledge and awareness fostered by digital literacy, which corresponds to the EA dimension [27]. Grounded in Bandura's self-efficacy theory [4], PSE is the user's belief in their ability to navigate and manage SVA privacy settings effectively. Finally, PPB is the behavioral expression of the capabilities and motivations of the other constructs. PPB represented the actions taken by youth to protect their personal information [28, 41]. In this way, PPB serves as the observable outcome of the dimensions of DOC, EA, and TT. This integrated model demonstrates that youth privacy behavior in SVA contexts is not driven by a single factor, but by a dynamic, multifaceted system.

**2.4 Prior Qualitative Work with Youth and SVAs**

This study is informed by our prior qualitative exploration of youth privacy with SVAs [11]. Through several focus groups with Canadian youth, we identified several key experiential themes: anxiety about always-on listening and covert recording, uncertainty regarding log visibility and data retention, confusion about privacy settings due to fragmented controls and policy overload, and the benefits of hands-free microtasks and entertainment.

These qualitative insights reinforced and refined the construct-specific survey items developed for this study. For example, the recurring concern about always-on listening validated the inclusion of PPR2 ("Worry about conversations being recorded without full awareness or consent"), while uncertainty regarding data retention affirmed the relevance of PPR4 ("Unease about the duration voice recordings are stored"). Similarly, confusion about privacy settings substantiated items such as ATT3 ("Feeling the apps are upfront in explaining data processing") and PSE1 ("Knowledge of how to access and adjust privacy settings"). The conveniences highlighted by youth in the focus groups support items like PPBf1 ("Extent to which voice-activated assistants save time and effort") and PPBf4 ("Appreciation for the apps' learning preferences to improve services"). Finally, reported protective behaviors, such as refusing certain permission and physically disabling microphones, corroborated items including PPB3 ("Refusal of certain features to maintain privacy") and PPB4 "Use of additional measures to protect data"). By aligning the survey instruments with these themes, this study ensures that the quantitative measures reflect actual lived experiences of young SVA users, thereby providing validation of the constructs. Table 1 summarizes all construct-level survey items.

Table 1: Constructs and Items

| Construct | Items |
| --- | --- |
| Perceived Privacy Risk (PPR) | PPR1: Concern about the amount of personal information collected by smart voice assistants.<br>PPR2: Worry about conversations being recorded without full awareness or consent.<br>PPR3: Belief that voice data could be accessed by unauthorized parties.<br>PPR4: Unease about the duration voice recordings are stored. |
| Perceived Privacy Benefits (PPBf) | PPBf1: Extent to which voice-activated assistants save time and effort.<br>PPBf2: Worth of sharing data for the personalized features offered.<br>PPBf3: Belief that benefits outweigh data collection worries.<br>PPBf4: Appreciation for the apps learning preferences to improve services. |
| Algorithmic Transparency and Trust (ATT) | ATT1: Understanding of the types of information collected and stored.<br>ATT2: Trust that manufacturers responsibly handle voice data.<br>ATT3: Feeling that apps are upfront in explaining data processing.<br>ATT4: Belief that apps provide fair and unbiased recommendations. |
| Privacy Self-Efficacy (PSE) | PSE1: Knowledge of how to access and adjust privacy settings.<br>PSE2: Capability to prevent apps from recording when undesired.<br>PSE3: Confidence to update permissions to increase data privacy.<br>PSE4: Belief in ability to effectively manage associated privacy risks. |
| Privacy-Protective Behavior (PPB) | PPB1: Frequency of reviewing or updating app permissions.<br>PPB2: Frequency of deleting voice search/activity history.<br>PPB3: Refusal of certain features to maintain privacy.<br>PPB4: Use of additional measures to protect data. |

**2.5 Gap and Positioning**

The current research landscape has produced valuable, yet fragmented, insights into youth privacy in AI-driven ecosystems. Prior work has established a high-level, multi-stakeholder model of privacy governance (the PEA-AI framework) and identified key experiential themes among youth through qualitative focus groups specific to SVAs [6, 11]. Although this work provides essential context, a critical gap remains between macro-level surveys and qualitative narratives. Specifically, the literature lacks a descriptive empirical study that quantifies youth perception and behaviors across core PEA-AI dimensions within the specific context of SVAs. Furthermore, a needed study must move beyond aggregate construct scores to uncover nuanced patterns in how youth respond to specific, scenario-based items related to SVA architectures. This granularity is essential for translating abstract principles into concrete design insights and targeted interventions. Finally, to be actionable, such research must directly connect these quantified lived experiences to system design and governance. Our study positions itself to fill this gap. By deploying a survey instrument whose constructs are grounded in prior qualitative work, we conduct a targeted descriptive analysis. Our goal is to establish a foundational quantitative profile of youth privacy perceptions in the SVA ecosystem.

**3 METHODS**

**3.1 Multi-Study Program Overview**

- Phase 1: PEA-AI stakeholder study
- Phase 2: SVA focus groups
- Phase 3 (this study): youth SVA survey.

**3.2 Participants and Recruitment**

Participants of this study are Canadian youth ages 16-24 who responded with at least one SVA use in the prior month. Participants were recruited through multiple channels, including flyers, emails, personal networks, LinkedIn, and through collaboration with several Canadian school districts and universities to better reach our targeted demographics of youth aged 16-24. A monetary incentive was offered to the first 500 survey respondents, with district-specific exceptions where required. A consent form was administered before starting the questionnaire. By submitting the consent form, participants indicated they understood the conditions of participation in the study as outlined in the consent form. We conducted online surveys through Microsoft Forms. Upon completing the questionnaire, participants were directed to a separate form to claim the incentive by providing their email address.

A total of 494 participants took part in the questionnaire. Responses were omitted that did not meet the demographic criteria (Canadian youth aged 16-24 with at least one SVA use in the prior month) or that contained insufficient data (≥ 20% missing responses). After data cleaning, 469 valid responses remained. Of those responses, n = 174 identified as female, n = 241 as male, and n = 15 as non-binary, n = 35 selected "prefer not to say", and n = 4 were missing/blank. The average participant's age was 18.65. 278 of the participants were High School students, while 183 had completed or were currently enrolled in Post-Secondary Education. Frequency of SVA use varied, with the largest portion reporting rare use of SVAs (n = 190), followed by daily use (n = 126), weekly use (n = 113), and finally monthly use (n = 38). Full demographic characteristics are highlighted in Table 2.

Table 2: Participant Demographic

| Characteristic | Categories | n | % |
|---|---|---|---|
| Gender | Blank/Missing | 4 | 0.9 |
| | Female | 174 | 37.1 |
| | Male | 241 | 51.4 |
| | Non-binary/Other | 15 | 3.2 |
| | Prefer not to say | 35 | 7.5 |
| Education Level | High School | 278 | 59.3 |
| | Post-Secondary | 183 | 39 |
| | Blank/Missing | 8 | 1.7 |
| Frequency of SVA use | Daily | 126 | 26.9 |
| | Monthly | 38 | 8.1 |
| | Rarely | 190 | 40.5 |
| | Weekly | 113 | 24.1 |
| | Blank/Missing | 2 | 0.4 |
| Age | | Mean (SD) = 18.65 (2.30) | |

## 3.3 Instrument Development

The development of the 20 items began with analyzing and adapting measures drawn from validated privacy and technology scales, including constructs related to privacy concern (IUIPC), privacy calculus (cost-benefit), self-efficacy, and transparency/trust. Wording was tailored using SVA focus-group themes, and items were mapped to the PEA-AI dimensions. PPR items are drawn from literature about privacy concerns and risk, but contextualized to SVA data processing. PPBf items were drawn from privacy calculus benefits, emphasizing perceived utility, convenience and personalization advantages. ATT items were drawn from transparency and trust measures, focusing on perceived clarity of data practices, system honesty, and institutional trust. PSE items were drawn from self-efficacy and privacy control literature, emphasizing perceived ability to manage privacy settings and disclosures. PPB items were drawn from privacy-protective behavior literature, focusing on disclosure management, configuration behavior, and device-level mitigation strategies. We briefly piloted the instrument with n = 6 empirical research specialists from Vancouver Island University and the University of Saskatchewan. Feedback was collected via follow-up discussion, and we made minor clarity edits, including rewording ambiguous items, standardizing terminology, and adjusting question order/formatting for readability.

## 3.4 Measures

This section contains definitions for each of the five constructs and one example item per construct. The full list of items is available in Table 1. Note that each item is measured on a 5-point Likert scale.

### 3.4.1 Perceived Privacy Risk (PPR)

Perceived privacy risk is defined as the extent to which youth feel vulnerable or at risk when using voice-activated AI apps.

PPR1: Concern about the amount of personal information collected by smart voice assistants.

### 3.4.2 Perceived Privacy Benefits (PPBf)

Perceived privacy benefits is defined as the perceived advantages or conveniences gained from using voice-activated AI apps that can offset privacy concerns.

PPBf1: Extent to which voice-activated assistants save time and effort.

### 3.4.3 Algorithmic Transparency and Trust (ATT)

Algorithmic transparency and trust is defined as the degree to which users believe that voice-activated AI developers are transparent about data practices, thereby fostering trust.

ATT1: Understanding of the types of information collected and stored.

### 3.4.4 Privacy Self-Efficacy (PSE)

Privacy self-efficacy is defined as users' confidence in their ability to identify, manage, and protect personal information when using voice-activated services.

PSE1: Knowledge of how to access and adjust privacy settings.

### 3.4.5 Privacy-Protective Behavior (PPB)

Privacy-protective behavior is defined as concrete actions taken by youth to safeguard their personal information and limit data collection in smart voice assistants.

PPB1: Frequency of viewing or updating app permissions.

## 3.5 Procedure and Ethics

Participants were provided with an informed consent form outlining the purpose of the study, data collection procedures, confidentiality expectations, and their right to withdraw at any time. After providing consent, participants completed the survey through Microsoft Forms, which consisted of questions tailored to the construct items. The survey required, on average, 6 minutes to complete. We measured responses to the items on a 5-point Likert scale ranging from 1 ("Strongly disagree") to 5 ("Strongly agree"). Higher scores indicate higher levels of the underlying constructs. We also collected data on control variables, including age, gender, educational level, and SVA usage frequency.

This study received ethics approval from the Vancouver Island University Research Ethics Board (VIU-REB). The approval reference number #103597 was given for consent forms, questionnaires, and behavioral/amendment forms.

## 3.6 Data Analysis

To evaluate the measurement quality of the constructs, we first conducted reliability and validity checks using PLS-SEM. Cronbach's alpha and composite reliability were computed in SmartPLS using our previously published model used in SVA privacy research [13].

Descriptive statistics were used to summarize overall participant perceptions at both the construct and item levels. At the construct level, we calculated means, standard deviations, and minimum/maximum values for all five constructs measured on a 1-5 Likert scale. At the item level, means and standard deviations were calculated for each of the twenty items. These descriptive analyses were conducted in R (v. 4.4.2).

Group comparisons examined whether perceptions varied across subgroups. We compared responses among participants with high vs. low PPB (top vs. bottom quartile), heavy vs light SVA users (daily/weekly vs monthly/rarely), age groups (16-18 vs 19-24) and across four gender categories. All group comparison analyses and heat map visualizations were conducted using R (v. 4.4.2).

# 4 RESULTS

## 4.1 Scale Reliability and Construct Descriptives

Our survey used a 5-point Likert scale to compare mean responses across five key constructs. Table 3 presents the descriptive statistics, providing initial insight into participants' perceptions. PPR had the highest mean score (M = 3.61; SD = 0.91), indicating that respondents generally reported a moderate to high level of concern about SVA data practices. The constructs of PPBf (M = 3.00; SD = 0.95), PSE (M = 2.97; SD = 0.83) and PPB (M = 3.03; SD = 0.77) all hovered near the scale midpoint. This suggests that participants are somewhat ambivalent about the benefits of SVA use outweighing the risks, feel moderately capable of managing their privacy, and occasionally engage in protective behaviors. In contrast, ATT had the lowest mean score (M = 2.52; SD = 0.72), indicating a relatively low understanding and trust in the companies and algorithms behind voice assistants. The standard deviations for all constructs showed a reasonable spread of responses.

We assessed convergent validity and internal consistency reliability for each of the constructs by calculating Average Variance Extracted (AVE) and composite reliability metrics, as shown in Table 4. Following established guidelines [18], AVE should exceed 0.50, indicating that 50% of the variance in the items is captured by the hypothesized construct. In our study, all constructs demonstrated good convergent validity, with AVE values exceeding the 0.50 threshold. Furthermore, both Composite Reliability (rho_c), Dillon Godlstein's rho (rho_a), and Cronbach's alpha exceeded the acceptable level of 0.70 for all constructs [18], confirming the measures' internal consistency reliability.

Table 3: Descriptive Statistics

| Construct | Mean (SD)    | ATT    | PPB    | PPBf   | PPR    | PSE |
|-----------|--------------|--------|--------|--------|--------|-----|
| ATT       | 2.52 (0.718) |        |        |        |        |     |
| PPB       | 3.03 (0.767) | 0.045  |        |        |        |     |
| PPBf      | 3.00 (0.949) | 0.414  | -0.115 |        |        |     |
| PPR       | 3.61 (0.909) | -0.276 | 0.307  | -0.334 |        |     |
| PSE       | 2.97 (0.825) | 0.449  | 0.308  | 0.278  | -0.128 |     |

Table 4: Construct Reliability and Validity

| Construct | AVE   | rho_c | rho_a | Cronbach's alpha |
|-----------|-------|-------|-------|------------------|
| ATT       | 0.529 | 0.818 | 0.712 | 0.714            |
| PPB       | 0.539 | 0.823 | 0.719 | 0.716            |
| PPBf      | 0.711 | 0.907 | 0.901 | 0.866            |
| PPR       | 0.734 | 0.917 | 0.907 | 0.880            |
| PSE       | 0.626 | 0.870 | 0.814 | 0.801            |

## 4.2 Item-Level Patterns by Construct

To gain a more granular understanding of participant perception and behaviors, we further analyzed the descriptive statistics for each individual survey item. The following subsections present the mean scores for each item, grouped by construct. The analysis reveals specific strengths, concerns, and behavioral gaps in our youth population. Table 5 presents all item-level statistics. Figure 1 shows the item-level distribution of means.

Table 5: Item-Level Descriptive Statistics

| Items | Mean | SD | Min | Max |
|---|---|---|---|---|
| ATT1 | 2.90 | 1.07 | 1 | 5 |
| ATT2 | 2.38 | 1.00 | 1 | 5 |
| ATT3 | 2.30 | 0.932 | 1 | 5 |
| ATT4 | 2.49 | 0.933 | 1 | 5 |
| PPB1 | 2.41 | 1.05 | 1 | 5 |
| PPB2 | 2.47 | 1.08 | 1 | 5 |
| PPB3 | 3.98 | 0.906 | 1 | 5 |
| PPB4 | 3.24 | 1.11 | 1 | 5 |
| PPBf1 | 3.14 | 1.19 | 1 | 5 |
| PPBf2 | 2.89 | 1.11 | 1 | 5 |
| PPBf3 | 2.80 | 1.14 | 1 | 5 |
| PPBf4 | 3.15 | 1.06 | 1 | 5 |
| PPR1 | 3.53 | 1.05 | 1 | 5 |
| PPR2 | 3.62 | 1.12 | 1 | 5 |
| PPR3 | 3.72 | 1.01 | 1 | 5 |
| PPR4 | 3.56 | 1.06 | 1 | 5 |
| PSE1 | 3.21 | 1.06 | 1 | 5 |
| PSE2 | 2.88 | 1.09 | 1 | 5 |
| PSE3 | 3.06 | 0.986 | 1 | 5 |
| PSE4 | 2.74 | 1.03 | 1 | 5 |

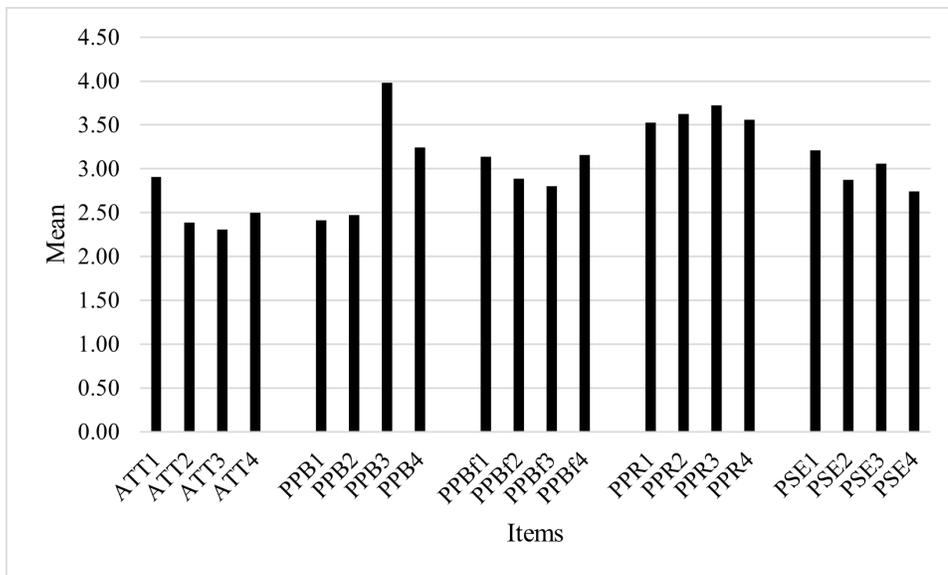

Figure 1: Distribution of means by survey item.

*4.2.1  PPR Items*

Responses to the PPR items reveal a consistent moderate to high level of concern across all measured risk factors, with means ranging from 3.53 to 3.72. The most salient risk was the belief that voice data could be accessed by unauthorized third parties (PPR3, M = 3.72), indicating this is perceived as a primary threat. Concerns about covert recording (PPR2, M = 3.62) and the long-term storage of data (PPR4, M = 3.56) were also prominent, while concerns about the volume of data collected (PPR1, M = 3.53) were slightly lower. This pattern aligns with the perceived risks dimension of the PEA-AI framework, highlighting that the security and secondary use of data is a critical point of user apprehension.

*4.2.2  PPBf Items*

A clear gap is noticeable between recognizing the conveniences and service improvements offered by voice assistants (PPBf1, M = 3.14; PPBf4, M = 3.15) and the broader evaluation of benefits versus privacy concerns (PPBf3, M = 2.80). While users acknowledge the utility and personalization of the applications, they are less convinced that these benefits justify the accompanying data collection, with the perceived worth of sharing data for personalization also scoring low (PPBf2, M = 2.89). The tension between recognized advantages and risk apprehension aligns with the core of the privacy paradox, and the enduring risk-benefit trade-off identified in prior work [2, 28, 42].

*4.2.3  ATT Items*

The ATT construct shows the lowest overall scores. The lowest-rated items reflect a fundamental lack of perceived transparency from companies (ATT3, M = 2.30) and low trust in manufacturers to handle data responsibly (ATT2, M = 2.38). Trust in algorithmic fairness (ATT4, M = 2.49) also scored poorly. Notably, participants' self-reported understanding of data practices (ATT1, M = 2.90) was rated slightly higher than other items, though still below the scale midpoint. This pattern highlights a critical transparency-trust tension, where relatively high self-assessed understanding is not sufficient to foster trust when users perceive companies as being opaque and untrustworthy in their data practices.

*4.2.4  PSE Items*

Participant's confidence in managing their privacy is moderate and uneven. While users feel most capable regarding knowledge of where to find settings (PSE1, M = 3.21) and updating permissions (PSE3, M = 3.06), their sense of efficacy drops for more dynamic actions. The weakest areas are the overall belief in their ability to manage risks (PSE4, M = 2.74) and the capability to prevent unwanted recording in real-time (PSE2, M = 2.88). This indicates that procedural knowledge does not fully translate into a sense of control over the technology's operation or confidence in mitigating risk, pointing to a need for better tools and education in the applied, situational aspect of privacy management.

*4.2.5  PPB items*

Reported engagement in PPB reveals an implementation gap between passive refusal and active management. The most common behavior was the refusal to use certain features (PPB3, M = 3.98), a relatively passive form of control. The use of additional protective measures (PPB4, M = 3.24) was less frequent. In contrast, active and ongoing technical management behaviors were the rarest: reviewing/updating permissions (PPB1, M = 2.41) and deleting voice history (PPB2, M = 2.47). This shows a clear behavioral pattern where avoidance is more common than active management, highlighting a disconnect between privacy concerns and applied privacy maintenance.

### 4.3 Patterns Across Usage, Behavior, and Demographic Groups

To understand how perceptions and behaviors vary among participants, we conducted group comparisons based on privacy-protective behavior engagement, usage frequency, and demographic factors. Mean heatmaps for comparisons are available in Appendix A.1 (Figures 2-5): Figure 2 compares high versus low PPB, Figure 3 reports SVA usage frequency, Figure 4 reports age groups, and Figure 5 reports gender identity. As a practical guide, mean gaps of ~0.30 points or more on the 1–5 scale are treated as practically meaningful for design/governance, with gaps ≥0.50 considered substantial; smaller gaps are interpreted as modest even when statistically significant.

#### 4.3.1 High vs Low Protective Behavior

Participants were divided into two groups based on their PPB scores: the High PPB group (scoring in the top 25%) and the Low PPB group (scoring in the bottom 25%). The mean scores for all constructs were compared across these groups, as shown in Table 6. The analysis revealed distinct profiles for high and low protective action. The most significant differences were found between PSE and PPR. Individuals in the High PPB group reported significantly greater confidence in their ability to manage privacy (M = 3.30 vs. 2.63; $p < 0.001$) and perceived significantly higher risks in SVA use (M = 3.94 vs. 3.31; $p < 0.001$). Notably, the High PPB group also reported slightly but significantly lower PPBf than the Low PPB group (M = 2.72 vs. 3.00; $p < 0.05$). Practically, the PSE (+0.67) and PPR (+0.63) gaps are substantial, supporting targeted interventions that make real-time control clearer and build self-efficacy, while the PPBf difference (−0.28) is comparatively modest. There was no significant difference in ATT between the groups. This suggests that engagement in protective behaviors is more strongly associated with a combination of high perceived risk and high self-efficacy than with levels of trust in companies.

Table 6: Group comparison by PPB levels.

|  | PPB Groups | | |
| --- | --- | --- | --- |
| Mean (SD) | Low PPB | High PPB | p |
| ATT | 2.43 (0.78) | 2.56 (0.76) | 0.197 |
| PPB | 2.11 (0.42) | 3.95 (0.44) | <0.001 |
| PPBf | 3.00 (0.98) | 2.72 (1.05) | 0.024 |
| PPR | 3.31 (1.03) | 3.94 (0.86) | <0.001 |
| PSE | 2.63 (0.89) | 3.30 (0.87) | <0.001 |

#### 4.3.2 Heavy vs Light SVA Usage

Participants were categorized as Heavy Users (reported daily or weekly SVA use) or Light Users (reported monthly or rare SVA use). The construct means across these groups are presented in Table 7. Heavy and Light SVA usage exhibited differing psychological profiles. As expected, Heavy Users perceived significantly greater benefits (PPBf) from technology (M = 3.54 vs. 2.43; $p < 0.001$) and reported slightly higher trust and transparency perceptions (ATT) (M = 2.69 vs. 2.35; $p < 0,001$). In contrast, Light Users perceived significantly higher risks (PPR) than Heavy Users (M = 3.72 vs. 3.50; $p < 0.01$).

From a practical standpoint, the PPBf gap (+1.11) is large enough to warrant usage-segmented design and messaging, whereas the ATT gap (+0.34) and PPR gap (−0.22) are smaller and should be framed as incremental, not transformative. There were no significant differences in their reported PPB or PSE. This pattern suggests that frequent use may be associated with a normalization of risk and a stronger focus on utility, while infrequent use is linked to higher risk perception.

Table 7: Group comparison by SVA usage.

|  | SVA Use | | |
| --- | --- | --- | --- |
| Mean (SD) | Light | Heavy | p |
| ATT | 2.35 (0.69) | 2.69 (0.71) | <0.001 |
| PPB | 3.09 (0.84) | 2.97 (0.69) | 0.086 |
| PPBf | 2.43 (0.84) | 3.54 (0.69) | <0.001 |
| PPR | 3.72 (0.94) | 3.50 (0.87) | 0.008 |
| PSE | 2.90 (0.86) | 3.03 (0.79) | 0.088 |

*4.3.3 Variations by Age and Gender Identity*

Comparisons by age group and gender revealed several notable patterns (see Table 8 and Table 9 for results). Regarding age, younger participants (16-18) reported significantly lower PPR and higher PPBf than older participants (19-24), but also higher ATT and lower PPB. This suggests a more utilitarian and less cautious stance among the younger users. In practical terms, the largest age difference is PPR (+0.41 for ages 19–24), suggesting age-tailored risk communication may be justified, while the remaining mean gaps (~0.19–0.25) are modest and better interpreted as fine-tuning signals rather than distinct "profiles." Notably, PSE did not significantly differ by age group.

Regarding gender, significant differences were found for ATT, PPBf, and PSE. Notably, Non-Binary participants reported the lowest levels of ATT and PSE, but the highest levels of PPR, indicating a particularly critical and wary perspective. Female participants reported significantly lower PSE than Male participants. These findings highlight that privacy perceptions and competencies are not uniform and can vary meaningfully across demographic groups. Practically, the non-binary group shows large differences in ATT (−0.7+) and PPBf (≈−1.0) and a moderate PSE gap (≈−0.6), but given the small subgroup size and non-significant PPR test, these patterns should be presented as descriptively important and in need of confirmation rather than definitive.

Table 8: Group comparison by age groups.

|  | Age Group | | |
| --- | --- | --- | --- |
| Mean (SD) | 16-18 | 19-24 | p |
| ATT | 2.60 (0.69) | 2.38 (0.77) | 0.002 |
| PPB | 2.97 (0.75) | 3.16 (0.78) | 0.012 |
| PPBf | 3.09 (0.93) | 2.84 (1.01) | 0.012 |
| PPR | 3.44 (0.93) | 3.85 (0.84) | <0.001 |
| PSE | 3.04 (0.78) | 2.92 (0.90) | 0.153 |

Table 9: Group comparison by gender identity.

|  | Gender | | | | |
| --- | --- | --- | --- | --- | --- |
| Mean (SD) | Male | Female | Non-Binary | Prefer Not to Say | p |
| ATT | 2.57 (0.75) | 2.54 (0.67) | 1.83 (0.54) | 2.38 (0.70) | 0.002 |
| PPB | 3.03 (0.78) | 2.96 (0.75) | 3.32 (0.80) | 3.24 (0.77) | 0.147 |
| PPBf | 3.08 (0.96) | 2.92 (0.85) | 2.10 (1.07) | 3.16 (1.02) | 0.001 |
| PPR | 3.59 (0.96) | 3.56 (0.87) | 4.22 (0.45) | 3.73 (0.83) | 0.094 |
| PSE | 3.08 (0.87) | 2.84 (0.74) | 2.50 (0.80) | 3.05 (0.84) | 0.008 |

**4.4 Mixed-Methods Integration: Connecting Numbers to Narratives**

The integration of our quantitative survey data with the qualitative focus group narratives from a prior, complementary study [11] provides a more nuanced and detailed understanding of the key privacy tensions youth navigate with SVAs. This mixed-methods approach reveals how abstract constructs manifest in lived experience, aligning with the core tensions outlined in the PEA-AI framework within the specific context of SVA use.

*4.4.1   The Risk-Benefit Tension*

Our survey data revealed a clear tension central to the privacy calculus. While PPR scored moderately high, PPBf hovered near the scale midpoint. Notably, heavy SVA users reported significantly lower risk perception and significantly higher perceived benefits than light users, suggesting that frequent use normalizes risk while amplifying perceived utility. This trade-off is heightened by youth narratives that detail a daily, situational calculus. Participants valued concrete, micro-task convenience, with one noting its utility, "It's really helpful when your hands are busy...," while simultaneously expressing persistent anxiety about data retention, "I worry about what happens to the recording of your voice and whether that's saved somewhere" [11]. This pattern connects with the PRB dimension of PEA-AI. Demonstrating how risk anxiety can be overridden by hands-busy convenience, and vice-versa, showing the privacy calculus to be dynamic and situational rather than static.

*4.4.2   The Transparency-Trust Tension*

The ATT construct had the lowest overall mean score in our survey. Item-level analysis showed particularly low ratings for understanding data practices and trusting companies, indicating a severe deficit in transparency that corrodes trust. These numbers reflect a lived experience of confusion and policy fatigue described in focus groups. Participants linked opaque privacy policies to passive acceptance of terms, stating, "I feel like it's in the terms and conditions, but you're not going to read that huge list," while another added, "I struggle with the settings app." [11]. These findings directly engage the TT dimension. Quantitative data confirms a systemic transparency failure, while the qualitative narratives specify the mechanisms of this failure: policy overload and hidden controls. This misalignment between corporate compliance and user experience is a core PEA-AI tension demonstrated in the SVA setting.

*4.4.3   The Control vs Behavior Gap*

Our analysis reveals a gap between concern and action. While PPR was high, PPB was only moderate, with active behaviors like deleting history being rare. Critically, PSE emerged as a key differentiator; those with high PPB scores reported significantly higher PSE than those with lower PPB. This pattern is explained by youth narratives of confusion and apathy. Participants reported not taking action because "I don't know where to push or what to do in settings," highlighting an EA gap [11]. Others expressed hopelessness, feeling that devices would record "no matter what," which suppresses action despite great concern [11]. Those who did act resorted to extreme physical controls, seeking verifiable DOC outside of opaque software settings, with one participant explaining, "I have my phone mic disconnected internally, then if I need to make a call or something, I just use Bluetooth." [11]. These insights highlight the interaction between two PEA-AI dimensions. The weak translation of risk perception into protective behavior is mediated by deficits in EA and eroded DOC. The qualitative data show that youth desire control but are inhibited by a lack of knowledge and clear feedback.

## 5 DISCUSSION

### 5.1 Summary of Key Patterns

A summary of the survey's core findings reveals several interconnected patterns that reveal the mechanisms of the privacy paradox, the well-documented gap between privacy concerns and protective action calculus [2, 5]. First, youth report moderate to high-risk perceptions (PPR M=3.61) alongside moderate perceived benefits (PPBf M=3.00), confirming that engagement with SVAs involves a continual privacy calculus. This trade-off is central to the paradox; the perceived value of hands-free convenience and personalization may outweigh stated concerns, leading to continued use despite acknowledged risks. The finding that heavy SVA users report significantly higher benefits and lower risk perception than light SVA users is consistent with two interpretations aligned with prior research. First, self-selection, where individuals with a calculus skewed toward benefits adopt frequent use [40], and normalization, where repeated use may acclimate users to risk cues and shift attention toward utility over time [23, 31].

Second, there is a noticeable deficit in transparency and trust (ATT M=2.52), the lowest-scoring construct. Youth report particularly low trust in companies and perceive them as not being upfront about data practices (ATT2, ATT3). This exists alongside a gap in privacy self-efficacy (PSE M=2.97), where youth know where settings are but feel less capable of preventing unwanted recording or managing overall risks. This combination of low trust in the platforms and low confidence in one's ability to act may contribute to the intention–action gap by reducing perceived agency, a core component of the privacy paradox.

Third, these perceptions translate into infrequent and passive protective behaviors (PPB M=3.03). The most common behavior is refusing to use certain features, while active management, such as deleting voice history or reviewing permissions, is less common. This indicates a significant intention-action gap. The subgroup analysis provides a key to understanding this: high protectors (High PPB group) are characterized by high risk-perception and high self-efficacy (PSE). This suggests that the privacy paradox is not just a risk-benefit analysis but is fundamentally shaped by an efficacy gap. When individuals lack the knowledge or perceived capability to act, inaction becomes the default despite their concerns. This dynamic mirrors findings in automation trust, where reliance develops over time, reducing the tension between the desire for control and the willingness to surrender it [25]. Finally, meaningful differences exist across subgroups. Demographic analyses further reveal that younger participants (16-18) and non-binary youth exhibit distinct profiles; descriptively, the latter reports higher risk and lower trust and efficacy, highlighting that privacy experiences are not uniform across groups. However, PPR differences did not reach statistical significance (p≈0.094) and the non-binary subgroup was small (n=15), so these patterns should be interpreted cautiously Lived experience of the individual user can deepen the control gap and intensify the tensions at the heart of the privacy paradox.

## 5.2 Negotiation of Privacy with SVA through PEA-AI

The patterns found in our analysis support the core tensions of the PEA-AI framework [6], demonstrating how youth negotiate privacy within the architecture of SVAs.

### 5.2.1 The Risk-Benefit Negotiation (PRB Dimension)

The moderate scores on both PPR and PPBf quantitatively capture the central trade-off youth make. The qualitative narratives explain this as a situational calculus. The "hands-busy" convenience of a quick timer or song choice can momentarily override anxiety about data retention [11]. The heavy vs. light SVA usage analysis further informs this tension. Use amplifies perceived utility (PPBf) and dampens perceived risk (PPR), effectively moving the privacy calculus toward acceptance through routine exposure. This dynamic negotiation of the PRB dimension shows that youth privacy decisions are not concrete choices, but fluid compromises shaped by immediate context and habitual use.

### 5.2.2 The Weakness of Transparency-Trust (TT Dimension)

The critically low ATT scores reveal a broken feedback loop between transparency and trust. Youth report not understanding data practices (low transparency), which directly fuels distrust in companies and algorithms (low trust). This quantitative finding is validated by qualitative accounts of "policy overload" and confusing, fragmented settings [11]. The result is a damaging PEA-AI tension, where platforms may technically provide information (e.g., in lengthy terms and conditions), but in a form that is inaccessible and fails to build trust. Consequently, youth disengage, accepting terms they do not understand, which undermines the foundation for meaningful consent or informed negotiation.

### 5.2.3 Control and Capability Gaps (DOC and EA Dimensions)

The gap between moderate PSE and infrequent PPB, and the critical role of PSE in differentiating high and low protectors, directly implicates the DOC and EA dimensions. Procedural knowledge (e.g., knowing where settings are) does not equate to a sense of control over technology's always-listening architecture. The qualitative data demonstrates this gap; youth express a desire for control but face an "efficacy gap", feeling that devices record "no matter what" or that they "don't know where to push" [11]. This highlights a failure in EA, which in turn hampers DOC. Those who do action control often resort to tangible, physical actions (e.g., disconnecting mics), seeking agency outside the opaque digital settings. This shows that for DOC to be realized, EA must provide not just knowledge, but actionable understanding.

## 5.3 Design Implications for SVA Platforms

The observed item-level pattern helps to provide a blueprint for intervention. Design must address the specific frictions identified, organized here into actionable principles.

*5.3.1    Principle 1: Make data flows visible and controllable*

This principle addresses two key privacy concerns: persistent concern about covert recording (PPR2) and the infrequency of log-clearing behavior (PPB2). To rebuild trust and enable meaningful control, SVA interfaces must make abstract data flows visible and actionable. This requires moving beyond subtle indicators, like changing LED colors, to implement clear, persistent mic status visualizations, through a clear, persistent transparency channel (e.g., an on-device screen where available, a companion-app dashboard, and unambiguous device cues such as standardized LED patterns supported by a simple "what the lights mean" legend), explicitly signaling when audio is being buffered, processed, or transmitted. Furthermore, control must be integrated more into the flow of use. Rather than burying data management deep within account dashboards, platforms should provide immediate, in-context options. Examples include a straightforward "Delete Last Interaction" option appearing after a command is completed, or monthly voice or notification prompts that guide the user to review and delete their voice history. This allows protective actions to integrate seamlessly into use rather than as a separate chore.

*5.3.2    Principle 2: Build skills through guided interaction*

To bridge the gap in PSE, particularly around real-time control (PSE2) and risk management (PSE4), design must shift from providing static information to fostering interactive learning. The platform should develop engaging, youth-friendly privacy onboarding that uses plain language and relatable examples to explain what data is collected, how it is used, and why it matters. Beyond initial setup, guided tutorials embedded within settings menus can walk users step-by-step through performing actions, such as changing a wake word or adjusting microphone permissions, therefore transforming procedural knowledge into practical applications. In addition, just-in-time explanations should be integrated into interactions. For example, before executing a command that accesses sensitive data, the assistant could say, "To give you directions, I'll use your current location. You can manage location settings here." This helps to build digital literacy and confidence, empowering users to make informed decisions about their privacy.

*5.3.3    Principle 3: Reduce friction for protective actions*

This principle targets the observed behavioral gap where passive feature refusal (PPB3) is common, but active privacy management (PPB1, PPB2) is less common. The goal should be to make protective action so easy that they become default rather than the exception. This can be achieved by designing one-touch privacy shortcuts, a few examples of this could be a "Privacy Clean-Up" button on a device's home screen or a voice command like "Delete my recent activity", that automates tasks. Companies should also move away from overbearing, all-or-nothing permission grants, switching instead to more granular, task-specific consents that give users more control without overwhelming them. For shared household devices, the system must support distinct user profiles with individualized privacy defaults. This allows young users to have their own auto-delete settings and permissions preferences, even on a family-use smart system. By minimizing the effort required to act, design can help to turn privacy concerns into consistent behavior.

*5.3.4    Principle 4: Adaptable autonomy across developmental stages*

This principle addresses the differences in privacy perceptions between younger (16-18) and older (19-24) youth, as well as the unique profile of non-binary participants. A one-size-fits-all privacy interface fails these varied needs. Instead, design should offer scaffolded autonomy, where the complexity of privacy controls is adaptable to the user.

For younger, less experienced users, interfaces could offer simplified, curated privacy models with progressive disclosure of settings. These models would align with education incentives by providing age-appropriate explanations. Furthermore, platforms should ensure that privacy explanations and threat models are inclusive of LGBTQIA+ experiences [30, 32]. In this sample, non-binary participants showed a descriptively higher PPR and significantly lower ATT and PSE; however, given the small subgroup size, these differences should be treated as a high-priority design signal that warrants confirmation in larger, purposively sampled studies. Intentionally including non-binary and transgender voices in multi-stakeholder design negotiations is essential for creating platforms that are genuinely safer and more trustworthy for all.

**5.4 Governance and Education Implications**

*5.4.1 For Regulators and Policymakers*

Aligning with PEA-AI's call for meaningful youth agency, regulations should move beyond notice-and-consent. Policies could mandate privacy-protective defaults for youth accounts (e.g., auto-delete timelines, mic-off-by-default for new devices) and set standards for age-appropriate, layered privacy notices that explain data use in clear, concise language.

*5.4.2 For Educators*

The PSE items can serve as a practical curriculum checklist for digital literacy. Educators can design activities around questions like: "Can students find their SVA voice logs?" "Can they explain what happens to their data after a command?" and "Can they change critical settings like the wake word or microphone access?" This moves privacy education from abstract principles to applied skills.

*5.4.3 For Multi-Stakeholder Negotiation*

The distinct patterns, like the efficacy gap among younger teens or the high-risk/low-trust profile of non-binary youth, provide a shared evidence base for co-designing better norms. Parents can use these insights to have more informed conversations, while platforms can develop more inclusive and developmentally sensitive family controls that balance protection with growing autonomy, addressing the PDS dimension.

The following matrix summarizes implications across stakeholder roles, mapped to PEA-AI dimensions to keep recommendations actionable and non-redundant.

Table 10: Stakeholder implications matrix (PEA-AI)

| Stakeholder lever | PRB (risk–benefit) | TT (transparency–trust) | DOC (control) | EA (agency/skills) |
|---|---|---|---|---|
| Platform design | Make benefits explicit without "dark patterns" | Layered, in-context explanations; plain language | In-flow controls (delete last interaction; one-touch privacy) | Guided walkthroughs; just-in-time prompts |
| Education | Privacy calculus exercises using SVA scenarios | Teach "policy fatigue" and how to verify claims | Practice logs, permissions, auto-delete settings | Skills checklist aligned to PSE items |
| Regulation / Governance | Youth-protective defaults (e.g., auto-delete timelines) | Standards for layered notices + auditability | Requirements for usable, accessible controls | Accountability for youth-facing UX evidence |

### 5.5 Contribution to PEA-AI and Privacy Theory

This study makes several contributions to the PEA-AI framework and youth privacy research. We successfully operationalize the PEA-AI dimensions in a concrete domain. By mapping our privacy constructs to those in the PEA-AI framework, we provide an empirical model of how the frameworks abstract tensions manifest in the SVA ecosystem. Furthermore, it reveals points for theoretical refinement. The critical mediating role of PSE suggests that the DOC and EA dimensions could be more explicitly linked in the framework to account for the psychological capability to exercise control. The starkly low levels of transparency and trust suggest TT may be a prerequisite for effective negotiations in other dimensions. Finally, it positions the survey instruments as a reusable tool for longitudinal and comparative research. It can be used to evaluate the efficiency of new design interventions (e.g., testing if a new dashboard improves PSE, ATT, and PPB), support cross-country comparisons of youth privacy perceptions, and be adapted for other voice or conversational AI contexts (e.g., AI chatbots). By connecting quantitative scales to qualitative narratives, it offers a robust methodology for capturing a nuanced, negotiated reality for privacy in AI-driven worlds.

## 6 LIMITATIONS AND FUTURE WORK

There are several limitations that should be considered. The Canadian online sample may limit generalizability to other regions and socio-technical contexts. Subgroup analyses, particularly for non-binary participants, should be interpreted cautiously due to small cell sizes, even when mean gaps appear large. Given multiple group comparisons across constructs and items, findings should be treated as exploratory and interpreted alongside practical magnitude, not p-values alone. Measures rely on self-reported behavior (PPB), which may diverge from observed settings used in real households. Finally, the instrument is tailored to SVA ecosystems; its portability to other conversational AI contexts should be empirically validated.

Future work includes longitudinal work to monitor changes in scores, intervention studies focused on privacy self-efficacy and behavioral change, comparative analyses across platforms and jurisdictions, and the integration of the instrument with more advanced modelling (SEM, tension indices, clusters), and validation studies that combine survey measures with observed or logged privacy actions (where ethically feasible), across multiple conversational-AI platforms.

## 7 CONCLUSION

The smart voice assistant (SVA) ecosystem functions as a practical test bed for PEA-AI, making visible how youth and emerging adults negotiate privacy under conditions of ambient sensing, shared household governance, and limited interface transparency. Using an operationalized survey instrument aligned to five core constructs—perceived privacy risk (PPR), perceived privacy benefit (PPBf), attitudinal transparency and trust (ATT), privacy self-efficacy (PSE), and privacy-protective behavior (PPB)—the study summarizes how privacy negotiation manifests in everyday use as a set of tensions between perceived value and perceived vulnerability, and between desired control and feasible action. The descriptive patterns indicate that privacy risk remains salient while perceived transparency and trust are comparatively low, and that protective behaviors tend to be uneven, with greater reliance on passive protections than on sustained, active management, consistent with an intention–action gap in youth privacy practice. The primary contribution of this work is the instrument and its mixed-methods integration, which connects survey evidence to prior qualitative insights to interpret observed tensions without over-claiming causal mechanisms. The results motivate privacy-by-design approaches that treat privacy as ongoing negotiation: clearer transparency channels (including screenless options such as companion-app dashboards and standardized device cues), in-the-moment controls that are easy to locate and verify, and capability-building supports that strengthen self-efficacy through just-in-time guidance. For platform governance and regulation, the findings reinforce the need to move beyond policy disclosure toward youth-centered transparency standards and usability-based evidence that consent and control mechanisms are meaningfully actionable. Future work should validate portability beyond SVAs, strengthen subgroup stability through larger purposive samples, and triangulate self-reported behaviors with ethically designed observational or log-based measures where feasible.

**ACKNOWLEDGMENTS**

This project has been funded by the Office of the Privacy Commissioner of Canada (OPC) through the Contributions Program Grant of the third author; the views expressed herein are those of the authors and do not necessarily reflect those of the OPC.

# A APPENDICES

## A.1 Item-Level Mean Heatmaps

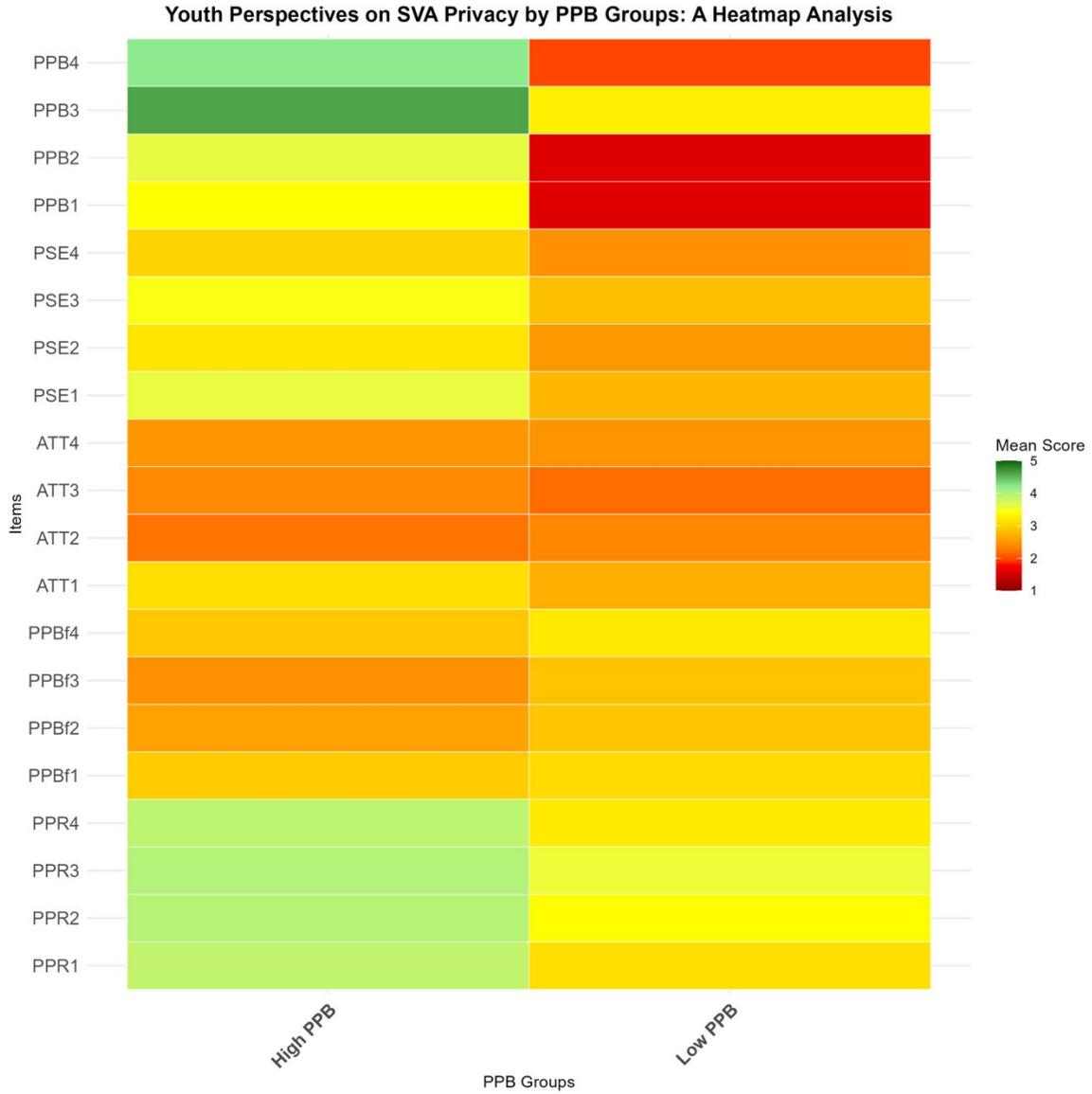

Figure 2: Heatmap of item-level means for high and low PPB.

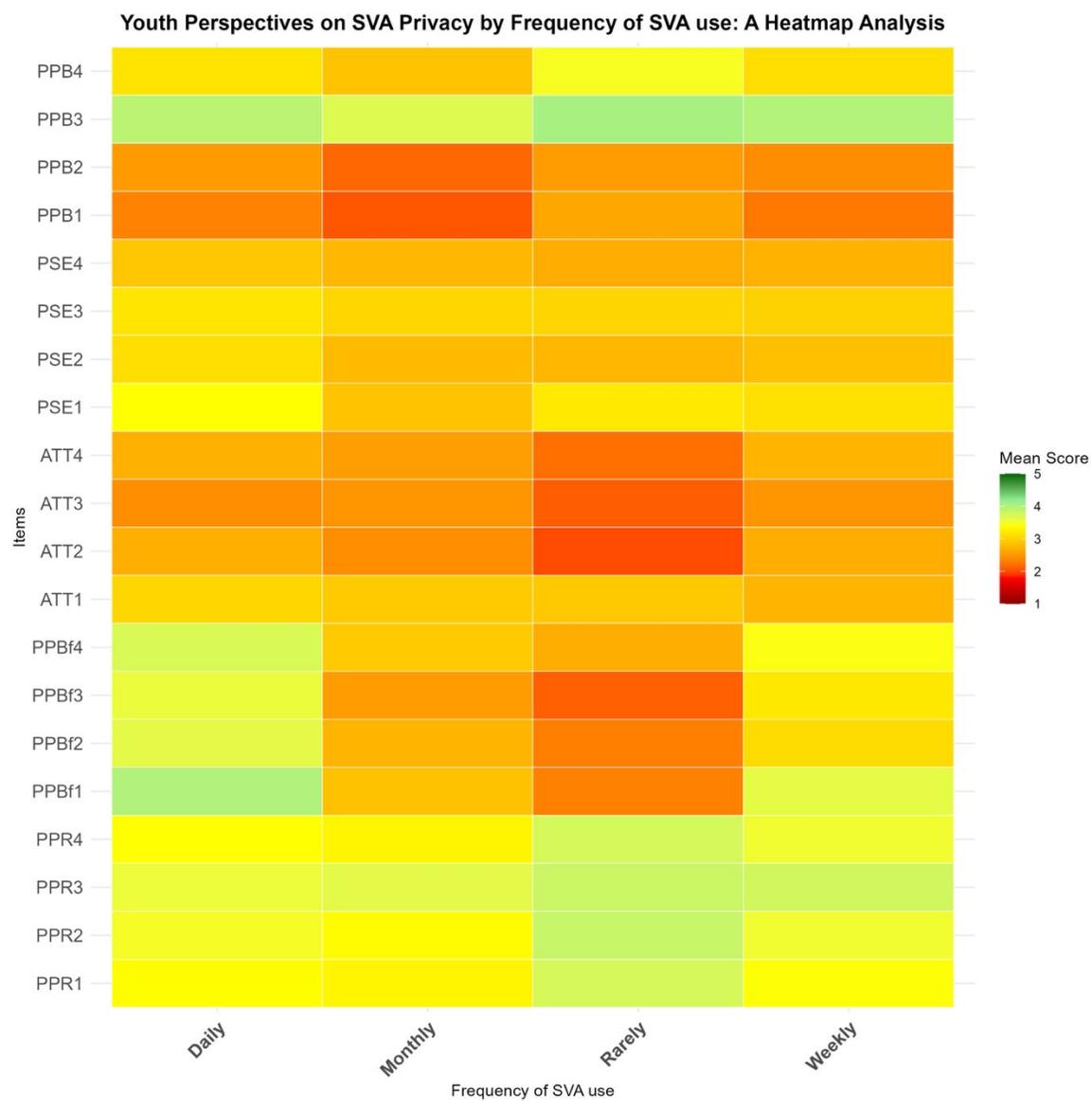

Figure 3: Heatmap of item-level means for SVA usage.

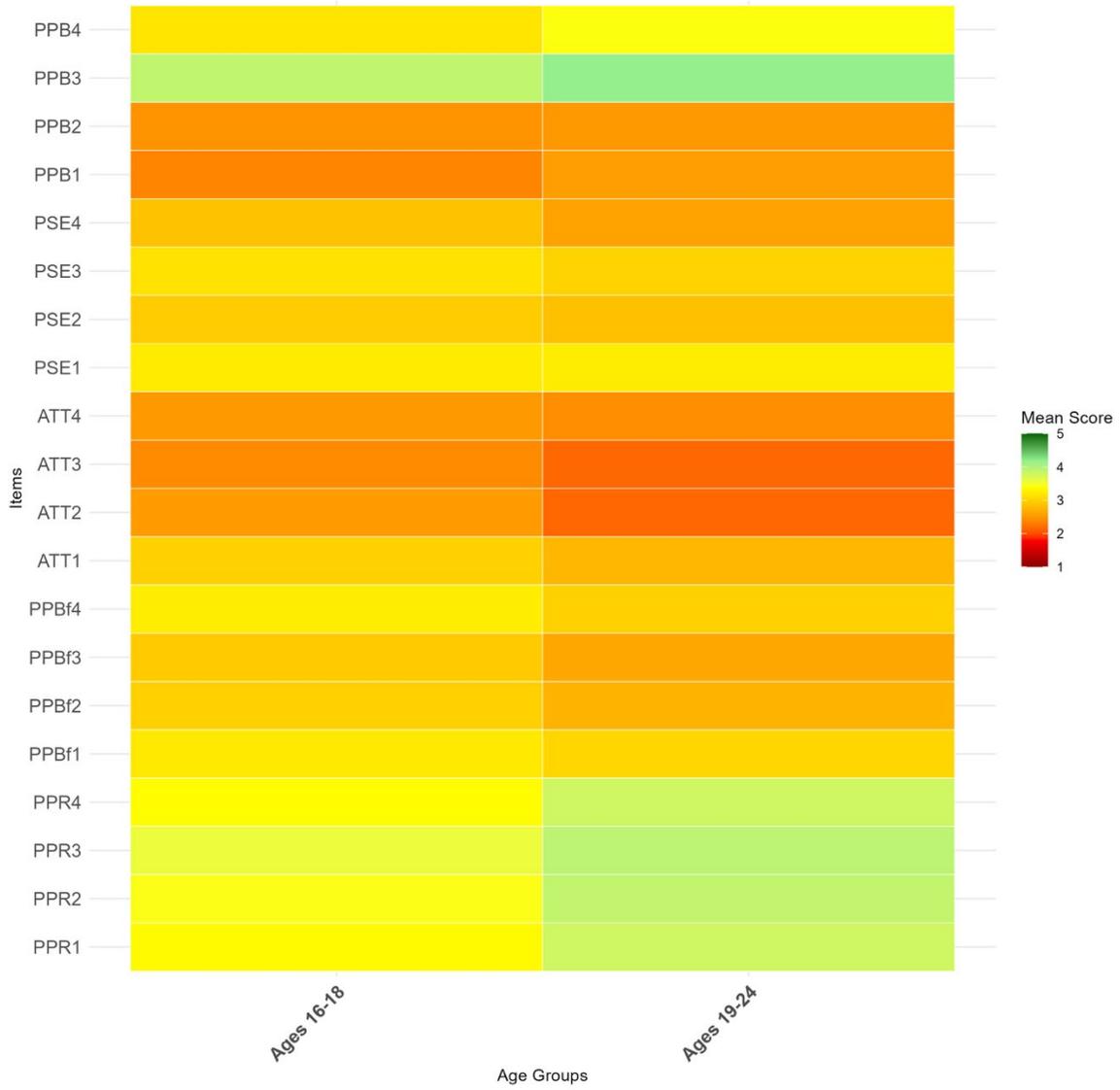

Figure 4: Heatmap of item-level means by age groups.

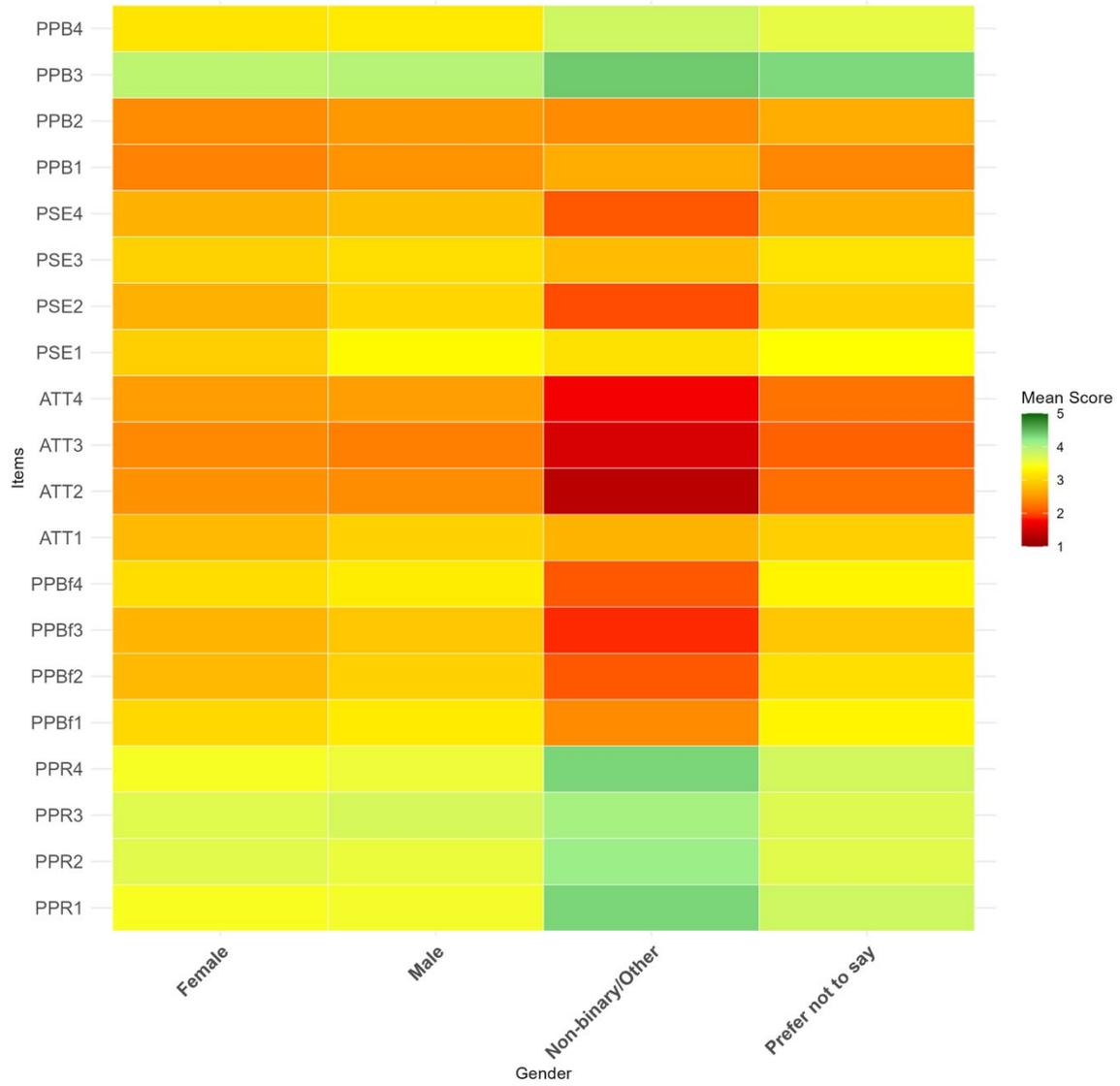

Figure 5: Heatmap of item-level means by gender identity.